\documentclass[twocolumn,english,aps,prb,twocolum,superscriptaddress,natbib,bibnotes,amsmath,amssymb,floatfix,groupedaddress,footinbib]{revtex4-2}
\pdfoutput=1
\usepackage[colorlinks=true,citecolor=blue,linkcolor=magenta]{hyperref}

\usepackage[markup=nocolor, authormarkupposition=left]{changes} 

\usepackage{soul}
\usepackage[utf8]{inputenc}
\usepackage[english]{babel}
\usepackage{amsmath,amsfonts,amssymb}
\usepackage[T1]{fontenc}
\usepackage{url}
\usepackage{amsmath}
\usepackage{siunitx}
\usepackage[version=4]{mhchem}
\usepackage{amsfonts}
\usepackage{amssymb}
\usepackage{epstopdf}
\usepackage{graphicx}
\graphicspath{{./Figures/}}

\usepackage{changes}
\usepackage{lineno}
\usepackage{titlesec}
\usepackage{gensymb} 

\begin{document}
\setlength\linenumbersep{4pt} 

\title[Article Title]{Self-Configuring Universal Multichannel and Multidimensional \\ Integrated Photonic Processing Engine}

\author{Zengqi Chen$^{1,2,\ddagger}$, Wu Zhou$^{1,2,\ddagger}$, Hao Chen$^{1,2,\ddagger}$, Kaihang Lu$^{1,2}$, Wenzhang Tian$^{1,2}$, Yiou Cui$^{1,2}$, Yuxiang Yin$^{1,2}$, Mingyuan Zhang$^{1,2}$, Xiaofu Pan$^{3}$, Jianqi Hu$^{3}$ $\&$ Yeyu Tong$^{1,2,*}$}

\affiliation{$^1$Microelectronic Thrust, The Hong Kong University of Science and Technology (Guangzhou), 511453, Guangzhou, Guangdong, China}
\affiliation{$^2$Guangdong-Macao Joint Laboratory for Modular Chip Design and Testing, The Hong Kong University of Science and Technology (Guangzhou), 511453, Guangzhou, Guangdong, China}
\affiliation{$^3$Department of Electrical and Electronic Engineering, The University of Hong Kong, Hong Kong, China}
\affiliation{{$^\ddagger$}These authors contributed equally to this work.} 
\affiliation{{$^*$}Corresponding authors: \href{mailto:yeyutong@hkust-gz.edu.cn}{yeyutong@hkust-gz.edu.cn}}

\maketitle

\noindent\textbf{\noindent Arbitrary manipulation of light across multiple physical dimensions is essential for harnessing its parallelism in fundamental research and advanced applications, such as optical interconnects, computing, imaging, sensing, and quantum networks. However, creating a universal device capable of arbitrary operations of multidimensional optical beams has been challenging, primarily due to their complex mutual interferences and dynamic transmission characteristics. In this study, we experimentally demonstrate a self-configuring integrated photonic processor designed for the arbitrary manipulations of multiple optical waves over their spatial and polarization dimensions. Despite the random nature of the input speckle, the photonic processor relies on an optical singular-value decomposition engine to sort all orthogonal input beams and implement arbitrary processing over both spatial and polarization dimensions precisely. Notably, the photonic processor can be self programmed in situ, enabling versatile functionalities such as beam shaping, optical switching, and reconfigurable optical add-drop multiplexing. Our findings advance the manipulation of multidimensional optical beams through a scalable, CMOS-compatible integration approach, paving the way for fully exploiting the parallelism of light in various applications.}


\section*{Introduction} 

Photons possess a multitude of exploitable dimensions, such as amplitude, phase, wavelength, polarization, and spatial structures \cite{forbes2021structured, wang2024integrated}. The full utilization of parallelism of light across various degrees of freedom is significant for both pioneering fundamental research and real-world applications. For instance, photonic multiplexing techniques can enable multidimensional entanglement in quantum networks \cite{zheng2023multichip}. Incorporating extra physical dimensions of light can expand data transfer capacity by orders for both free-space \cite{zhao2015capacity} and fiber optics communications \cite{puttnam2021space, rademacher2021peta}. Future technological roadmaps can thus be reshaped, ranging from remote satellite communications to meter-level inter-chip connections \cite{yang2022multi} and even micro-scale optical interconnects on chips \cite{sun2025edge}. These have been intensively investigated recently for distributed high-performance computing systems \cite{hosseini20218} and artificial intelligence accelerators \cite{ shen2017deep, zhou2022photonic, chen2022iterative}. Moreover, multidimensional optics can also substantially enhance resolution and accuracy in imaging and microscopy applications \cite{zhang2022large}. Although the mathematical framework and design concept have been proposed \cite{miller2013self, miller2013complicated, miller2013reconfigurable}, a universally applicable multichannel and multidimensional design for arbitrary beam manipulation has yet to be experimentally realized.

When dealing with $M$ distinct input beams, conventional photonic elements such as lenses, mirrors, plates, and filters can only realize a limited subset of the possible functions. Identifying efficient elements working simultaneously for multiple input beams over various dimensions necessitates extensive computational effort and performance compromises \cite{liu2012highly}. Furthermore, general physical operation on optical waves must work beyond fixed input wavefront to include those that are also time-varying. This consideration is critical due to the complex mutual interferences among multiple multidimensional input beams and their dynamic transmission characteristics within the physical media through which light propagates. Various channel-dependent mechanisms, such as phase distortion, polarization rotation, attenuation, and dispersion, can significantly affect the wavefront of the beams. For example, the free-space transmission of light can be significantly impacted by stochastic, time-varying atmospheric turbulence \cite{martinez2024self,zaminga2025optical}. Even in guided-wave optics, such as optical fibers with silica cladding, their optical transmission matrix can function as a `black box', undergoing substantial changes due to environmental perturbations, including twisting, sharp bends, temperature fluctuations, and variations in mechanical stress \cite{rademacher2021peta, lu2024empowering}. Due to the complex mutual interferences and the different propagation dynamics, the resultant chaotic and time-varying speckle poses a considerable challenge for direct processing using fixed physical designs lacking adaptability. Hence, multidimensional optical beams behave like uncontrolled forces, making them challenging to manage in the optical domain. As a result, the practical applications of multidimensional optics are significantly constrained.

Arbitrary processing of light is particularly well suited to microscopic integrated systems \cite{fontaine2012space, yang2022multi, wan2024multidimensional}. The recent rapid advancement of photonic integrated circuits has revolutionized light-matter interactions \cite{shekhar2024roadmapping, bogaerts2020programmable}. Planar photonic waveguides provide a scalable platform for reconfigurable manipulation of light across multiple degrees of freedom. The nanowire waveguide structure can be engineered to exhibit strong birefringence, facilitating the on-chip separation of various spatial and polarization modes into different waveguides. This allows arbitrary multi-beam processing within photonic integrated circuits \cite{ribeiro2016demonstration}. In recent years, reconfigurable photonic processors consisting of Mach-Zehnder interferometers (MZIs) or micro-ring resonators (MRRs) have been extensively investigated for processing of multimode optical signals, including signal decomposition processors \cite{sharma2025universal, milanizadeh2022separating, annoni2017unscrambling, zaminga2025optical,lu2024empowering, zhang2024system, tang2021ten, yi2024unmixing}, multimode structured light generators \cite{butow2024generating, wu2023chip, zhou2025adaptive, zhao2025all}, and high-dimensional beam analyzers \cite{butow2023photonic, butow2022spatially, sharma2025universal, gu2025disordered}. Through self configuration, the optimal spatial mode set can be determined without prior knowledge of the free space system \cite{seyedinnavadeh2024determining, miller_self-configuring_2013}. Nonetheless, arbitrary operations on multiple and multidimensional beams with random input nature still lack crucial features, including functions like beam shaping, reconfigurable optical switching, and add-drop management across both spatial and polarization dimensions.

In this work, we first experimentally demonstrate the feasibility of a universal integrated photonic processor capable of arbitrary parallel processing of multiple and multidimensional optical waves. By utilizing integrated multidimensional antennas with optical singular value decomposition (SVD) mesh implemented through reconfigurable MZIs, our system effectively decomposes, processes, and reassembles arbitrary spatial and polarization beams. This approach enables the identification of eigenwaves from random speckle inputs, facilitates arbitrary physical operations for each orthogonal beam channel, and ensures efficient remapping to different output dimensions. Hence, a reconfigurable and arbitrary relationship between the inputs and the outputs can be realized. Notably, functionality of the integrated photonic processor can be programmed in situ with a simple self configuration process. We experimentally validate the programmable functionalities of our integrated photonic processor, including arbitrary beam structure shaping, optical switching, and reconfigurable optical add-drop multiplexing. Our proposed approach is both compact and scalable, enabling efficient, adaptive, and reconfigurable management of a multidimensional optical system directly within the optical domain. This capability unlocks the parallelism of light for a wide range of advanced photonic applications in the future.

\begin{figure*}
  \includegraphics[width=0.95\linewidth]{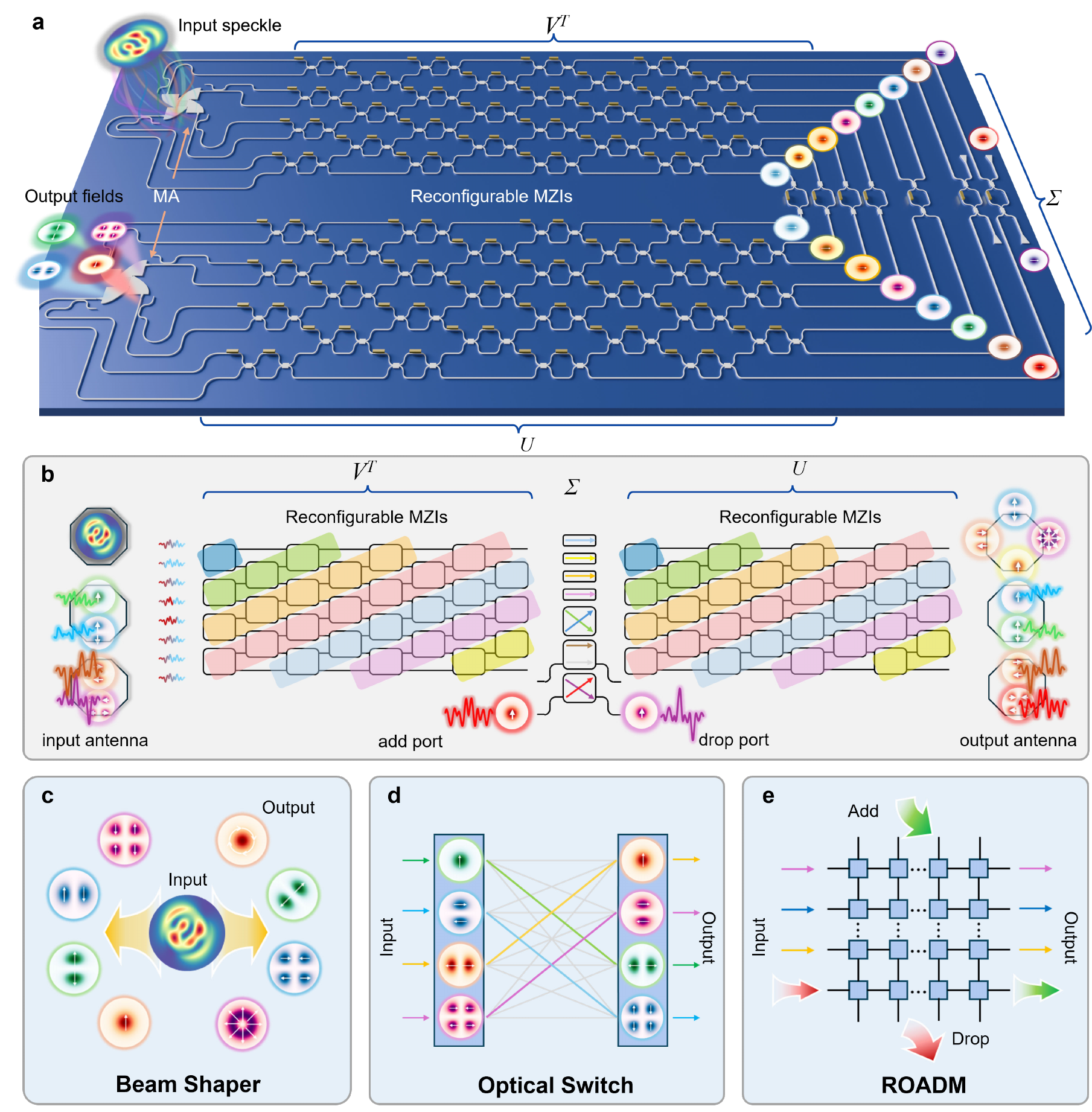}
  \caption{\textbf{Schematic, functionalities, and working principle of the integrated multichannel and multidimensional photonic processing engine.} \textbf{a} Schematic of the integrated photonic processor developed for multidimensional optical system. The photonic processor is designed on silicon photonics platform, with input/output multidimensional antennas (MAs) for capturing/launching various spatial and polarization beams, and reconfigurable Mach-Zehnder interferometers (MZIs) photonic singular-value decomposition (SVD) mesh. The complex input speckle is formed due to random mutual interferences of multiple input beams. It is decomposed into parallel quasi transverse electric (TE) modes on chip by the optical mesh \(V^{T}\). Arbitrary processing is thus implemented through \(\Sigma\), followed by remapping to different dimensions of output beams by the optical mesh \(U\). \textbf{b} Operation principle of the integrated photonic processor. The octagon represents the input and output MAs. The rectangle represents the tuneable MZIs. The left and right three pairs of input and output depict the operation of three different functions shown in \textbf{c}-\textbf{e}. The colored MZI diagonal lines illustrate the eight-dimensional unitary transformation employed in the optical decomposition and recombination process. \textbf{c} Illustration of beam shaper, where complex input speckle can be arbitrary transformed into the desired output spatial and polarization of optical wave. \textbf{d} Illustration of optical switch, which enables arbitrary inter-channel switching of optical signals among orthogonal beams in the optical domain. \textbf{e} Illustration of reconfigurable optical add-drop multiplexer (ROADM), which can be used to add, drop, pass and redirect optical signals with the external in order to customize the multidimensional optical system.}
  \label{fig:boat1}
\end{figure*}

\section*{Operation Principle}\label{sec2}

The proposed integrated silicon photonic processor is schematically depicted in Figure 1a. It is composed of multidimensional input and output optical antennas connected with a reconfigurable MZI-based SVD optical mesh. These optical antennas are identical diffraction grating couplers \cite{zhou2024ultra} designed to support eight orthogonal spatial and polarization modes with a constant unitary transmission matrix $A$ . Operation principles, design details, and simulation results of the optical antenna can be found in the Supplementary Note S1. With multiple beams $\sum\limits_{i=1}^{8}|\phi_{Ii}\rangle$ consisting of different spatial and polarization distributions shine into the input antenna, complex speckle pattern can be formed due to their random interferences and differing transmission characteristics. Nevertheless, they will be essentially decoupled and split into eight fundamental quasi transverse-electric (TE) modes on chip with a random amplitude and phase distribution via the input antenna. After the above non-selective down-conversion process, we utilize eight-dimensional interferometric SVD factorization mesh based on Clements scheme following the formula \cite{clements2016optimal}:

\begin{equation}
P= U\Sigma V^{T},
\label{eq1}
\end{equation}

\noindent where \(P\) represents a reconfigurable and arbitrary optical transmission matrix enabled by the MZI mesh. Thermal-optical phase shifters are employed to dynamically configure the two unitary transformation matrix \(U\) and \(V^{T}\). Hence, the proposed photonic processor relates multiple inputs $\sum\limits_{i=1}^{8}|\phi_{Ii}\rangle$ and outputs $\sum\limits_{i=1}^{8}|\phi_{Oi}\rangle$ through

\begin{equation}
\sum\limits_{i=1}^{8}|\phi_{Oi}\rangle=A^\dagger U\Sigma V^{T}A\sum\limits_{i=1}^{8}|\phi_{Ii}\rangle.
\label{eq2}
\end{equation}

Multiple input beams $\sum\limits_{i=1}^{8}|\phi_{Ii}\rangle$ are factorized dynamically by on-chip SVD through a forward transmission process. In response to the desired operations, the diagonal matrix \(\Sigma\) containing all eigenwaves can be manipulated by other integrated electro-optic components, such as modulators or switches, before being remapped to the desired output dimensions using the second unitary matrix \(U\) and the output antenna. The mathematical space of the integrated photonic processor is primarily limited by the complete orthonormal sets of spatial and polarization modes allowed by the integrated antenna and the reconfigurable mesh. 

Figure 1b depicts the operation principle of photonic processor. Three pairs of input and output corresponds to the three demonstrated functions in this work: beam shaper, optical switch, and reconfigurable optical add-drop multiplexer (ROADM) as summarized by Figures 1c-1e. Correspondingly, the diagonal matrix design can adopt 1$\times$1 MZIs, 2$\times$2 MZIs, or incorporate additional add and drop ports. The proposed beam shaper effectively decouples all the orthogonal channels from input complex speckle patterns and transforms them into arbitrary desired spatial or polarization modes at the output antenna, as depicted by Figure 1c and the first row of Figure 1b. As an optical switch, the photonic processor facilitates arbitrary routing of optical signals across up to eight orthogonal spatial and polarization channels without requiring photoelectric conversion. The second row of Figure 1b illustrates the exchange of carrier waves of two optical signals. Finally, by incorporating additional add and drop ports into the diagonal matrix on-chip, as shown in the last row of Figure 1b, add-drop operations can be flexibly configured without disrupting other concurrent optical channels, as illustrated in Figure 1e. It is important to note that additional linear and nonlinear operations on optical waves can be achieved in the future by integrating various optical components or gain amplifiers \cite{liu2019high} into the diagonal matrix, utilizing the recently developed hybrid or heterogeneous III-V materials on silicon \cite{kaur2021hybrid}.

Multidimensional optical antenna with design schematic depicted by Figure 2a is one of the key components in the optical processing system. For clarity, we elucidate its operating principle when employed as an optical emitter. The multidimensional optical antenna employs a shared diffraction grating region to project the $\text{TE}_{0}$ and $\text{TE}_{1}$ modes from eight different waveguides vertically out of the photonic chip plane. Between the diffraction region and eight strip waveguides, four pairs of $\text{TE}_{0}-\text{TE}_{1}$ mode multiplexer and spot-size converters (SSCs) are employed, which are based on asymmetrical directional coupler (ADC) \cite{ding2013chip, wang2013silicon} and parabolic-mirror collimator \cite{xu2023compact}. By injecting four pairs of $\text{TE}_{0}$ and $\text{TE}_{1}$ modes into the same diffraction grating region from four perpendicular directions, their coherent beating and relative amplitude and phase offset can determine the launched beam structure. The orthogonal basis of the multidimensional antenna can be defined by the eight linearly polarized (LP) modes depicted in Figure S1e. Figure 2b illustrates various polarization states of the launched quasi Gaussian beam, including horizontal linear polarization (HLP), vertical linear polarization (VLP),  $-45 \degree$ linear polarization ($-45 \degree$LP) and $+45 \degree$ linear polarization ($+45 \degree$LP), left-circularly polarization (LCP), and right-circularly polarization (RCP). Via injection of $\text{TE}_{1}$ modes into the optical antenna, other high dimensional modes including $\text{LP}_{11ax/y}$, $\text{LP}_{11bx/y}$, $-45 \degree$ rotated $\text{LP}_{11a/b}$, right- and left-circularly polarized Laguerre-Gaussian (LG) beam $\text{LG}_{01-RCP}$ and $\text{LG}_{01-LCP}$ modes can be selectively launched as illustrated by Figure 2c. Leveraging the reciprocity of light, the optical antenna can efficiently capture a disordered speckle resulted from arbitrary superposition of the eight orthogonal basic spatial and polarization modes. 

\section*{Results}\label{sec3}
Figures 2d-2f depict the microscopic images of the wire-bonded photonic processor implemented on silicon-on-insulator (SOI) platform after being fabricated by commercial silicon photonic foundry through a multi-project wafer (MPW) run. Due to architecture symmetry of the unitary matrices \(U\) and \(V^{T}\), we utilize two identified silicon photonic chips with a footprint of $\text{1.15 mm} \times \text{6.83 mm}$ containing the same MZI-based optical mesh. The two photonic chips were bridged with single-mode optical fiber array in our experiment. In our proof-of-concept demonstration, the diagonal matrix \(\Sigma\) processing is relying on the external fiber-based components, which can be upgraded to integrated waveguide components on silicon in the future. Zoomed-in views of the multidimensional optical antenna and tunable MZIs through thermal heaters are depicted in Figures 2e-2f. The diffraction grating region is consisted of 70-nm shallowly etch hole array with a size of $13.39 \, \mu m \times 13.39 \, \mu m$.

To validate the selective launching capability of various beam structures through the multidimensional antenna, we present the measured optical polarization and intensity profiles of an individual testing antenna using a polarization analyzer, an infrared camera with a $10\times$ imaging lens. The measurement setup is illustrated in the supplementary Figure S2a. Figures 2g and 2h present the measured fundamental Poincaré sphere and Bloch sphere for visualizing the generated polarization and spatial beam, including Gaussian beam with different linear or circular polarizations, and Hermite-Gaussian (HG) modes. All the beams are generated using the testing antenna with appropriate setting of the inputs as illustrated in Figures 2b and 2c. Supplementary Figure S2c-S2j present the experimental emitting efficiency spectra and intensity profiles of the optical antenna for eight orthogonal basic modes using few-mode optical fibers. 

\begin{figure*}
    \centering
    \includegraphics[width=0.95\linewidth]{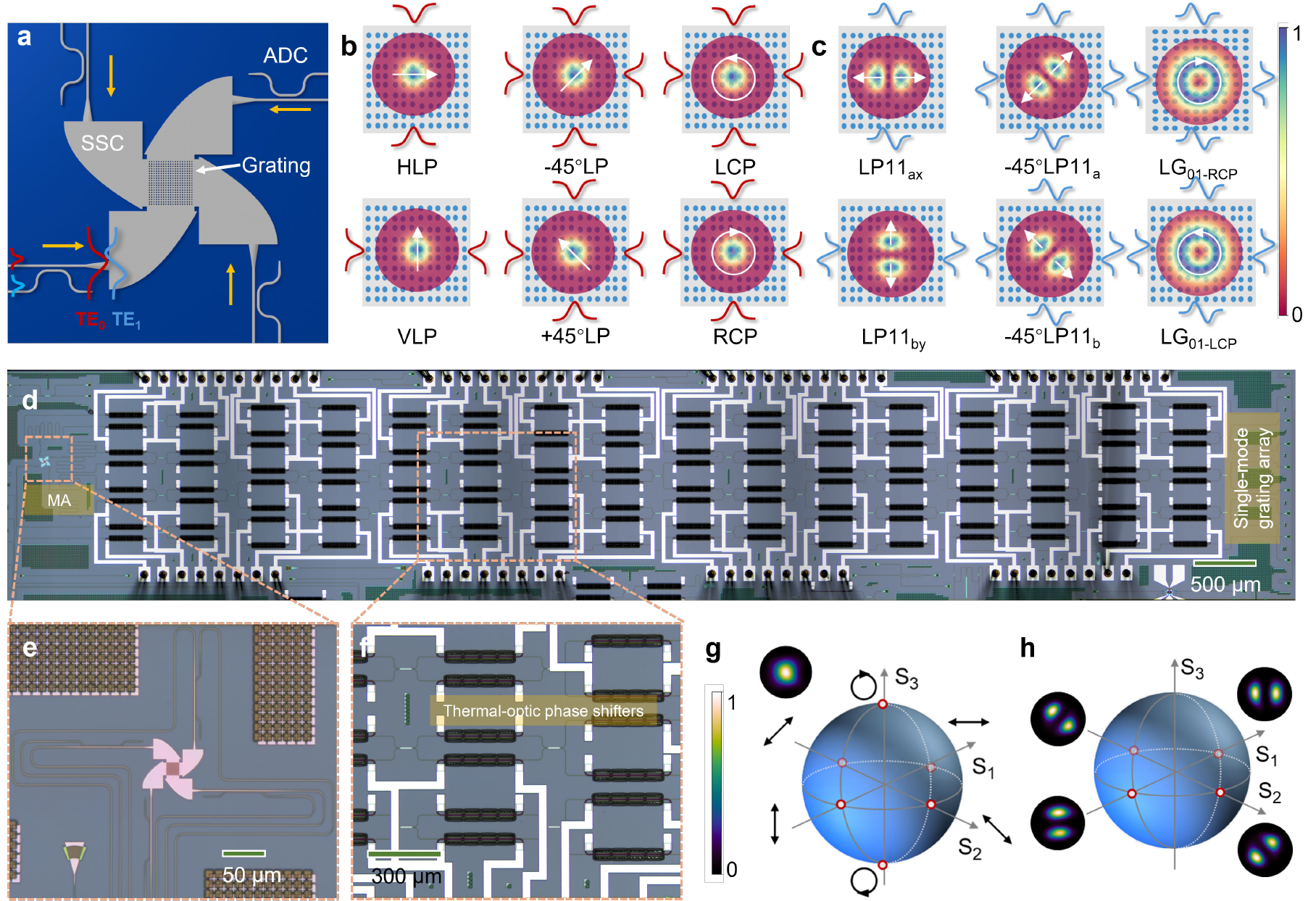}
    \caption{\textbf{Multidimensional optical antenna and silicon photonic processor.} \textbf{a} Operation schematic of multidimensional optical antenna, including asymmetrical directional coupler (ADC) based multiplexer, spot size converter (SSC), and diffraction grating coupler. \textbf{b} Launching schematic of various states of polarization, including horizontal linear polarization (HLP), vertical linear polarization (VLP), $-45 \degree$ and $+45 \degree$ linear polarization (denoted as $-45 \degree$LP and $+45 \degree$LP), as well as left-circularly polarized (LCP) and right-circularly polarized (RCP) Gaussian beams.
     \textbf{c} Launching schematic of various spatial and polarization modes, including $\text{LP}_{11ax}$, $\text{LP}_{11by}$, $-45 \degree$ rotated $\text{LP}_{11ax}$ and $\text{LP}_{11by}$, right- and left-circularly polarized Laguerre-Gaussian (LG) beam $\text{LG}_{01-RCP}$ and $\text{LG}_{01-LCP}$. \textbf{d} Integrated silicon photonic processor comprising of multidimensional antenna,  reconfigurable $8\times8$ photonic mesh based on 28 thermal-optical Mach-Zehnder interferometers (MZIs), and single-mode grating coupler array. Two identical silicon photonic dies containing the same MZI-based optical mesh bridged with single-mode optical fiber arrays are utilized in our demonstration. The diagonal matrix \(\Sigma\) processing is relying on the external fiber-based optoelectronic components, which can be upgraded to integrated waveguide components on silicon in the future. \textbf{e} $\&$ \textbf{f} Zoomed-in view of the multidimensional antenna and heater-based MZI. \textbf{g} $\&$ \textbf{h} Fundamental Poincaré sphere and Bloch sphere for visualizing different polarization and spatial beams \cite{forbes2021structured}, including Gaussian, and Hermite-Gaussian (HG) modes generated by the individual testing multidimensional antenna on silicon.}
    \label{fig:2}
\end{figure*}

\begin{figure*}
    \centering
    \includegraphics[width=0.95\linewidth]{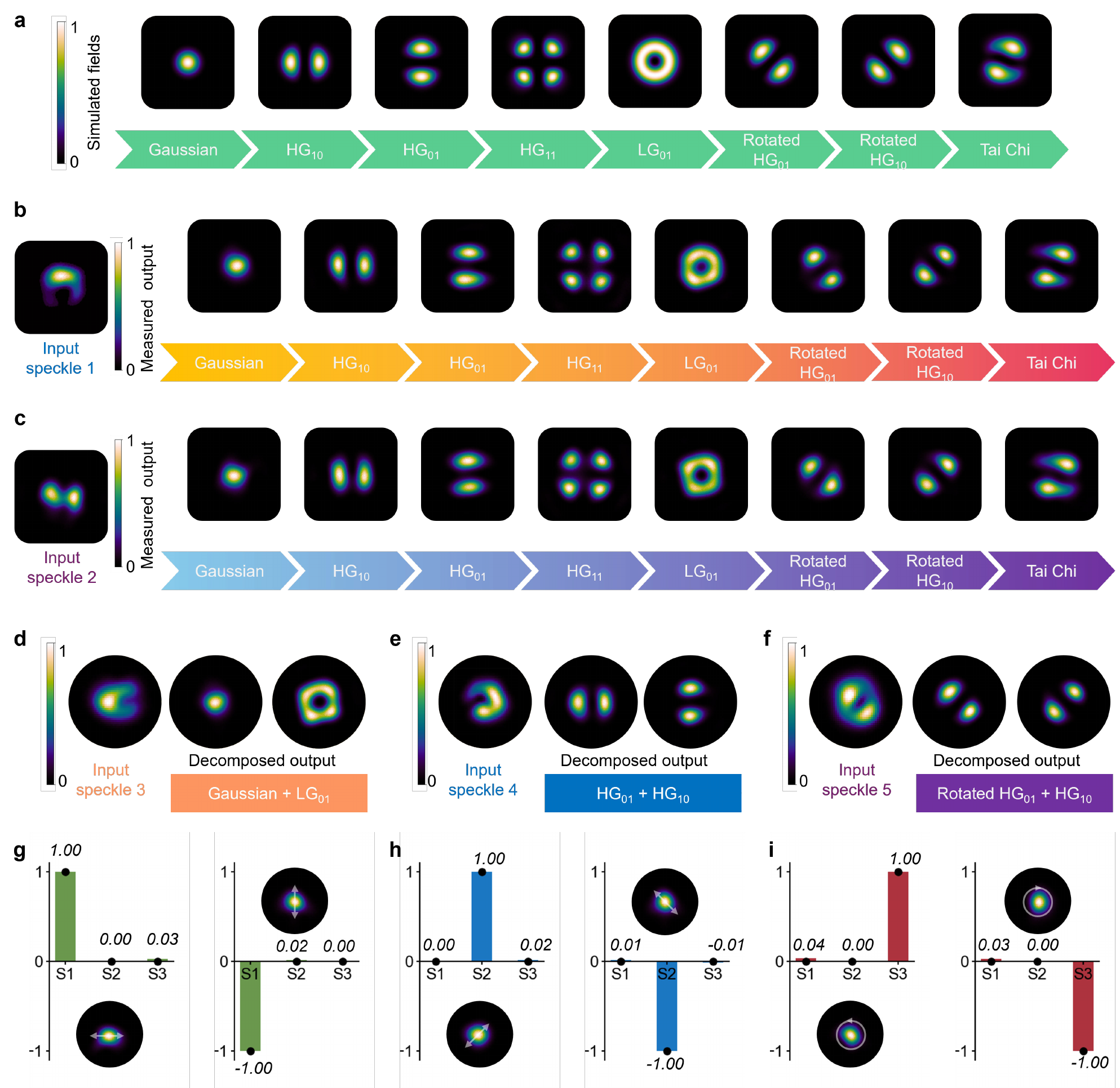}
    \caption{\textbf{Beam shaping with the integrated photonic processor.} \textbf{a} Target beam intensity distribution obtained from simulation, including Gaussian beam, Hermite-Gaussian (HG) beam $\text{HG}_{01}$, $\text{HG}_{10}$, $\text{HG}_{11}$, Laguerre-Gaussian (LG) beam $\text{LG}_{01}$, rotated $\text{HG}_{01}$ and $\text{HG}_{10}$, and non-eigenmode Tai Chi. The overlap integral between the measured intensity distribution with target intensity distribution is utilized as the feedback for configuring the photonic processor. \textbf{b} A random input speckle pattern is sent into antenna of the photonic processor via a few-mode optical fiber. By configuring the photonic processor with a multichannel programmable power supply, the output speckle pattern can be selectively shaped towards the target intensity distributions depicted in \textbf{a}. A 10x imaging lens and an infrared camera are used to capture the beam intensity profile. \textbf{c} Another random input speckle pattern by deliberately twisting the optical fiber is launched into the photonic processor. All the target beam profiles can be successfully reproduced. \textbf{d}-\textbf{f} Three random input speckle patterns containing two orthogonal optical channels are processed by the photonic processor, which can be successfully decomposed and remapped into three different pairs of orthogonal output beams, including Gaussian with $\text{LG}_{01}$, $\text{HG}_{01}$ with $\text{HG}_{10}$, rotated $\text{HG}_{01}$ with $\text{HG}_{10}$. \textbf{g}-\textbf{i} Measured output beam intensity profiles and Stokes parameters using a polarization analyzer. Three random input speckle patterns containing two orthogonal optical channels are processed by the photonic processor, which can be successfully decomposed and remapped into three different pairs of orthogonal polarized Gaussian beams.}
    \label{fig:3}
\end{figure*}

\subsection*{Beam Shaper}

In contrast to the above beam-generation approach, which depends on predetermined amplitudes and phase offsets of the optical signals directed into testing antenna, the beam shaping experiment presented here employed a dynamic and disordered speckle input, potentially consisting of a varying number of orthogonal channels. This non-selective decoupling leads to a random distribution of optical signals across the eight different single-mode waveguides subsequent to the input optical antenna. Consequently, the received random optical signal undergoes initial processing by the unitary matrix \(V^{T}\) to extract the eigenwaves. These eigenwaves are subsequently directed to each output waveguides of \(V^{T}\) with self configuration process before being processed by the diagonal matrix \(\Sigma\). The outputs from the diagonal mesh undergo transformation through the operator (\(U\)), projecting them onto designated elements of the output vector space. This process facilitates arbitrary spatial and polarization conversion for multiple input beams simultaneously. In our experiment, the on-chip SVD optical mesh was adaptively configured using a multichannel programmable power supply. The driving currents of the silicon photonic processor were optimized through the particle-swarm optimization (PSO) algorithm \cite{kennedy1995particle}. Beam decomposition was implemented through the unitary matrix \(V^{T}\) based on a simple power maximum process with external optical power meters. At the output optical antenna, a \text{$10\times$} objective lens and an infrared camera were employed to capture the optical intensity profile. The polarization state can be determined using a polarization analyzer. The overlap integral of the beam profile or the state of polarization are utilized as the feedback for controlling the operator (\(U\)). The detailed self configuration process of the integrated photonic processor is illustrated in the Supplementary Note S2. Detailed experimental setup for beam shaping can be found from the Supplementary Note S3A.

Figure 3a depicts the target beam intensity obtained from simulation, including Gaussian beam, Hermite-Gaussian (HG) beam $\text{HG}_{01}$, $\text{HG}_{10}$, $\text{HG}_{11}$, Laguerre-Gaussian (LG) beam $\text{LG}_{01}$, rotated $\text{HG}_{01}$ and $\text{HG}_{10}$, and non-eigenmode Tai Chi beam. In our experiment, a disordered input speckle pattern was sent to the photonic chip via a few-mode optical fiber with mode selective photonic lantern \cite{fontaine2012geometric, leon2014mode}. Owing to dynamic polarization rotation and intermodal coupling in the optical fiber, the speckle pattern of the output beam exhibits significant variability in response to fiber twisting and bending. Figure 3b illustrates the input speckle pattern directed into the silicon photonic chip. By adaptively configuring the silicon photonic processor, this random input speckle can be decomposed and subsequently reshaped to any of the desired beam structure emitted from the output antenna. The measured intensity profiles for all targeted beam patterns were obtained with high fidelity, as shown in Figure 3b. A second random input speckle was sent to the photonic processor by intentionally twisting the input optical fiber. The input speckle was shown in Figure 3c. This new input speckle can also be effectively reshaped to any of the desired output intensity profiles, yielding pure and clean intensity distributions.

The integrated photonic processor can also deal with multiple input beams. In our experiment, two concurrent orthogonal beams were launched via a fiber photonic lantern with a dynamic and random speckle pattern sent to the photonic chip, as presented in Figure 3d. The input speckle was first decomposed into two orthogonal single-mode channels, which were utilized to generate Gaussian and $\text{LG}_{01}$ mode at the output antenna respectively. Supplementary Video 1 demonstrates time-dependent fluctuations in the experimentally measured intensity profile from the output antenna. These dynamics arise from coherent interference between co-propagating fundamental Gaussian and $\text{LG}_{01}$ modes, which have a random phase offset due the unavoidable phase drifts in optical fibers before the input ports. Figures 3e and 3f present the experimental results with another two random input speckle patterns, which were successfully decomposed and reshaped to two orthogonal output modes, $\text{HG}_{01}$ with $\text{HG}_{10}$, rotated $\text{HG}_{01}$ with $\text{HG}_{01}$. Notably, Supplementary Videos 2 and 3 demonstrate distinctly different spatiotemporal evolution of the intensity profiles compared to Video 1, showcasing persistent central null regions. This phenomenon arises from the zero intensities present at the center of the output $\text{HG}_{01}$ and $\text{HG}_{01}$ modes.

Apart from shaping intensity profile, the photonic processor also facilitates simultaneous manipulation of both polarization and spatial structure, with appropriate feedback on the spatial and polarization status of the output beam. In our experiment, the output beam was equally split and sent to a polarization analyzer and an infrared camera respectively for precise beam structure monitoring, as shown by the setup in Supplementary Figure S5b. Figures 3g–3i present the measured intensity profiles and Stokes parameters for three pairs of orthogonally polarized Gaussian beams reshaped from two scrambled input beams. After a forward transmission process through the adaptively configured integrated photonic processor, the output beam intensity profile can be accurately shaped to quai Gaussian distribution with arbitrary polarization status, yielding HLP paired with VLP, $–45\degree$LP with $+45\degree$LP, and LCP with RCP. The deviation of the normalized Stokes parameters from their ideal target values remains within $\pm 0.04$ for all reported states. These results demonstrate precise controlling ability of the integrated photonic processor over both spatial and polarization states of multiple input beams. Due to the limitations of our polarization analyzer, which is unable to operate beyond the fundamental mode, precise polarization tracking of the higher-order modes was not implemented in our experiment. 

\begin{figure*}
    \centering
    \includegraphics[width=0.95\linewidth]{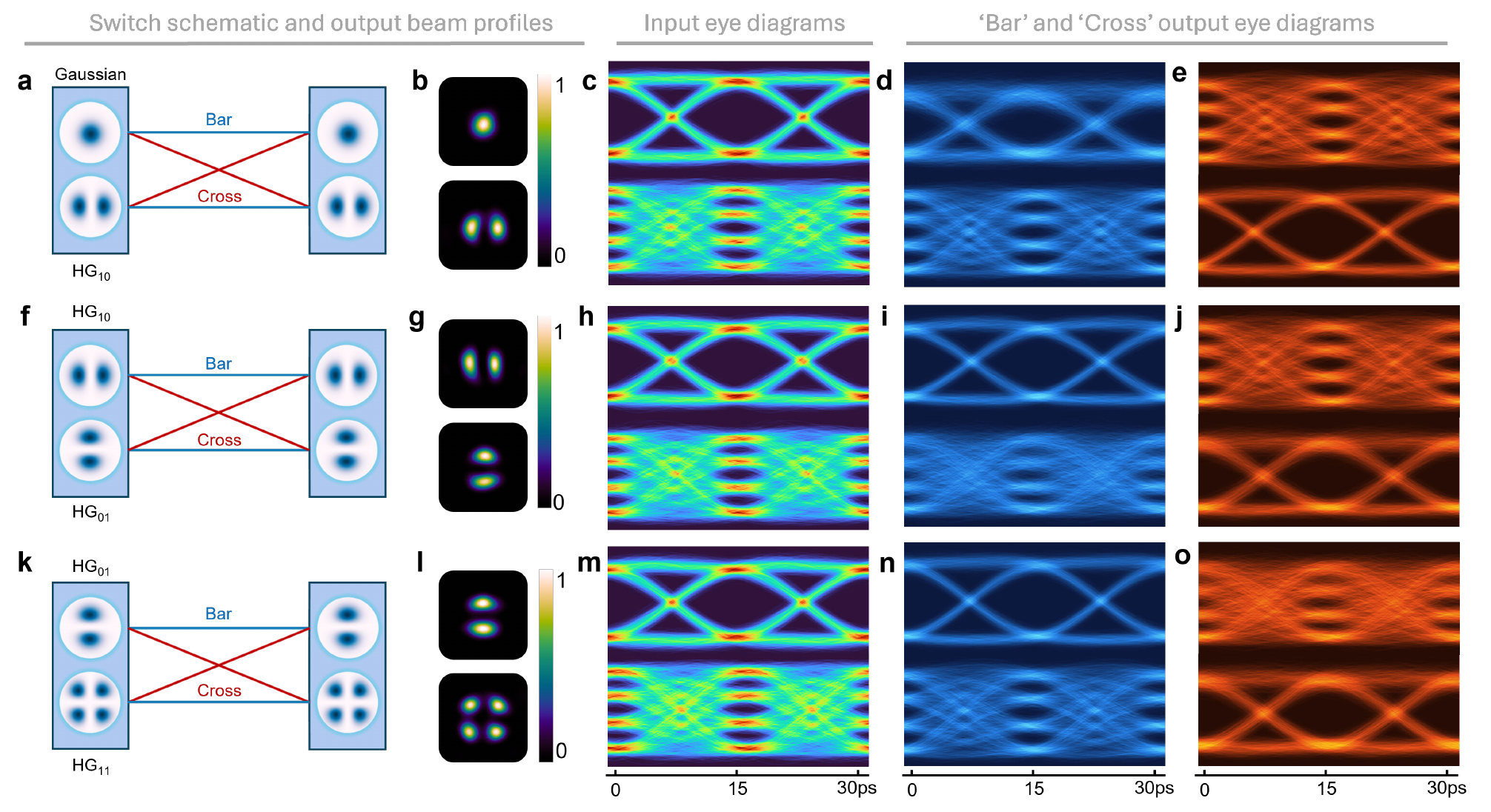}
    \caption{\textbf{Optical switch with the integrated photonic processor.} \textbf{a-c} Switch schematics, output beam intensity profiles, and input reference eye diagrams when working with Gaussian and $\text{HG}_{10}$ modes. The 64-Gbaud non-return-to-zero on-off keying (NRZ-OOK) and four-level pulse amplitude-modulation (PAM-4) signals were encoded to the input Gaussian and $\text{HG}_{10}$ modes via external electro-optical modulators and mode-selective fiber photonic lantern, respectively. Following processing through the photonic integrated circuits, the NRZ-OOK and PAM-4 optical signals are successfully routed to different output carrier modes $\text{HG}_{10}$ and Gaussian, or vice versa. \textbf{d-e} Measured output eye diagrams, where the blue-graded eye diagrams correspond to `Bar' state and the red-graded eye diagrams represent `Cross' state. \textbf{f-o} Switch schematics, output beam intensity profiles, input reference eye diagrams, and output eye diagrams when working for other two concurrent modes: $\text{HG}_{01}$ with $\text{HG}_{10}$, and $\text{HG}_{01}$ with $\text{HG}_{11}$. The carrier modes of the high-speed optical signals can be dynamically switched.}
    \label{fig:4}
\end{figure*}

\subsection*{Optical Switch}

In a multidimensional optical system, reconfigurable optical switches are essential for the dynamic routing of optical signals encoded on carrier waves with different spatial and polarization modes, thereby enhancing transmission efficiency and flexibility without necessitating additional electro-optical conversion. In our experiment, we employed two external Mach-Zehnder modulators alongside a fiber-based photonic lantern to generate two concurrent 64-Gbaud high-speed optical signals. Distinct modulation formats are employed for the two input beams, including non-return-to-zero on-off keying (NRZ-OOK) and four-level pulse-amplitude modulation (PAM-4). Subsequently, the optical signals were directed to the photonic chip for decomposition, crossbar switching, and remapping into the desired output orthogonal beam carriers. Detailed experimental setup can be found from the Supplementary Note S3B.

Figure 4a depicts the crossbar switching schematic for the Gaussian and $\text{HG}_{10}$ modes,  which are utilized to transmit 64-Gbaud NRZ-OOK and PAM-4 signals, respectively. The optical mesh was first configured to accommodate two concurrent input beams and map them selectively into the output Gaussian and $\text{HG}_{10}$ modes, as evidenced by the measured output intensity profiles presented in Figure 4b. By adaptively reconfiguring the photonic mesh, the carrier wave for high-speed NRZ-OOK signals can be directed to either the output Gaussian or $\text{HG}_{10}$ modes. Figures 4c-4e present the measured input reference eye diagrams, output eye diagrams when the switch is working at the 'Bar' or 'Cross' state. Notably, the high-speed signals transmitted via the Gaussian mode demonstrate a transition from NRZ-OOK to PAM-4. The input and output eye diagrams for the $\text{HG}_{10}$ mode are also presented, illustrating a modulation format switch from PAM-4 back to NRZ-OOK. The photonic processor chip is capable of supporting other orthogonal basic vectors, such as $\text{HG}_{01}$ paired with $\text{HG}_{10}$, and $\text{HG}_{01}$ paired with $\text{HG}_{11}$. Figures 4f to 4o provide illustrations of the switching schematics, the output beam profiles, and the input reference and the output eye diagrams when the switch was working at different states.

\subsection*{ROADM}

Different from the ROADM utilized in a wavelength-division multiplexing system, the demonstrated ROADM here is to add, drop, or pass through specific spatial and polarization modes in a multidimensional optical system \cite{miller_self-configuring_2013}. ROADM facilitates both communication and switching capabilities with external systems. Dynamic routing and management of optical channels can thus be allowed beyond the wavelength domain via our integrated photonic processor. In our experiment, since the photonic processor are designed with a spatial and polarization dimensionality of eight, Figure 5a depicts the add-drop schematic working for eight basic orthogonal spatial and polarization modes, including $\text{LP}_{01x/y}$, $\text{LP}_{11ax/y}$, $\text{LP}_{11bx/y}$, and $\text{LP}_{21ax/y}$. Detailed experimental setup is illustrated in the Supplementary Note S3C. Eight orthogonal spatial and polarization modes were generated via the external fiber photonic lantern and directed to the input optical antenna. Leveraging the on-chip photonic mesh, these spatial and polarization modes were converted into eight distinct fundamental TE modes before being selectively dropped. Due to the reciprocity of light, the drop ports can also function as add ports, enabling the multiplexing of all eight fundamental TE modes into a multidimensional optical system. Figure 5b shows a normalized transmission bar chart for various spatial and polarization input modes decomposed and dropped by the photonic processor. By accurately configuring the on-chip photonic mesh, the measured channel crosstalk at all drop ports remains below -16 dB. Figure 5c illustrates an alternative dropping scheme that maps the input beams to different drop ports, achieving a similar channel crosstalk performance. To illustrate the self-configuration process when the integrated optical mesh was adaptively controlled, Figures 5d and 5e present the measured optical transmission evolution of the eight drop ports when a single input beam was selectively dropped through different ports. Channel crosstalk serves as feedback to fine-tune the driving currents using the PSO algorithm during the self-configuration process. On average, achieving a stable and optimized channel crosstalk level took approximately one minute. Once the integrated photonic processor is correctly configured for selective channel dropping, the output drop ports can serve as the add ports when the input beam is injected oppositely. Figure 5f shows the clean measured intensity profiles when the add ports were utilized to excite various spatial modes.

To assess the performance of the processor with multiple beams, Figure 5g illustrates the add-drop-pass schematic, demonstrating the simultaneous addition, dropping, or passing of high-speed optical channels. In our experiment, orthogonal beams were selectively added or dropped through single-mode ports. Figures 5h and 5i present the corresponding normalized transmission bar charts for different selective dropping states of the two input beams, both achieving channel crosstalk levels of less than -25 dB. Additionally, high-speed optical eye diagrams were also measured in our experiment to verify the capability of the photonic processor for communication applications. Figure 5j displays the measured reference eye diagrams for the two concurrent input modes using 64-Gbaud NRZ-OOK and PAM-4 signals. When an incorrect photonic mesh configuration was employed, the optical signals at the two drop ports became mixed, as illustrated in Figure 5k. The photonic processor can be adaptively configured to allow the optical signals to pass through, yielding clean eye diagrams as shown in Figure 5l. Alternatively, the input signals can also be selectively dropped, with corresponding clean eye diagrams depicted in Figure 5m. Finally, optical signals can be added, with eye diagrams measured from the output antenna presented in Figure 5n.

\begin{figure*}
    \centering
    \includegraphics[width=0.95\linewidth]{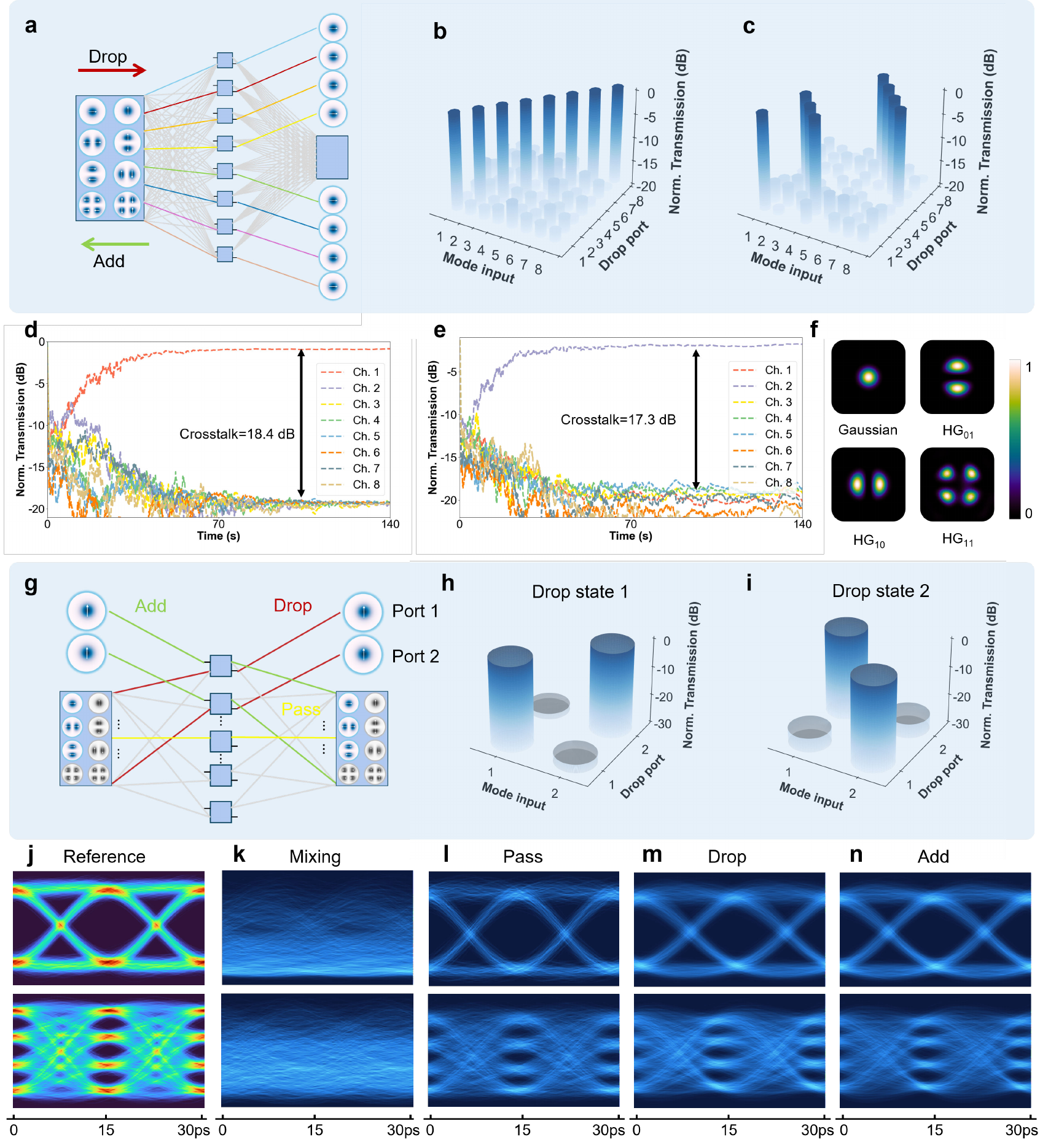}
    \caption{\textbf{Reconfigurable optical add-drop multiplexer (ROADM) with the integrated photonic processor.} \textbf{a} add-drop schematic illustrating that multidimensional optical waves can be selectively added or dropped via the integrated photonic processor. \textbf{b} Normalized transmission bar chart of the ROADM illustrating the channel-to-channel transmission and crosstalk. \textbf{c} Different drop configuration scheme via adaptively tuning the integrated photonic processor. \textbf{d} $\&$ \textbf{e} Optical transmission evolutionary diagrams of the eight drop ports during the self-configuration process when one of the input beam is selectively dropped. \textbf{f} Measured output beam intensity profile when orthogonal spatial modes are selectively added by the multidimensional antenna. \textbf{g} Add-drop-pass schematic and experiment with concurrent high-speed optical signals. \textbf{h} $\&$ \textbf{i} Optical transmission and crosstalk bar charts when two concurrent input beams were selectively dropped to different ports. \textbf{j-n} Measured eye diagrams of the ROADM by the oscilloscope, including input reference eye diagrams where two concurrent modes are encoded with different modulation formats, NRZ-OOK and PAM-4. Closed eye diagrams shown in \textbf{k} result from random and coherent beating between the mixed channels due to incorrect photonic processor configuration. However, once the photonic processor is correctly configured, high-speed optical signals can effectively pass through the ROADM, be selectively dropped, or added via add-drop ports as shown in \textbf{l-n}, respectively.
    }
    \label{fig:5}
\end{figure*}

\section*{Discussion}\label{sec3}

In summary, we proposed and demonstrated a self-configuring silicon photonic processor capable of physically manipulating multichannel optical beams in spatial and polarization dimensions, despite of their random mutual interferences and dynamic transmission characteristics. Using programmable waveguide meshes controlled via real-time feedback loops, our system adaptively works as a reconfigurable beam shaper, optical switch, or add-drop multiplexer. Crucially, this approach enables arbitrary programming of input-output beam relationships while leveraging the inherent scalability and cost-effectiveness of CMOS-compatible silicon photonics. Future integration with other advanced optoelectronic components, such as high-speed modulators \cite{lu2025whispering} or amplifiers \cite{liu2019high}, may further extend to the nonlinear operations of multidimensional optical waves. Our work establishes foundational hardware platform for multidimensional light processing applicable to next-generation optical interconnects, computing, imaging, and sensing applications.

Beyond space and polarization control, our proposed processing approach can extend to other physical dimensions of light, such as time and frequency, by incorporating additional on-chip photonic elements and making use of the available optical bandwidth of the entire system. Supplementary Figure S2 confirms the wide operational bandwidth of the multidimensional optical antenna. To further expand spatial and polarization channels, multidimensional antenna is one of the key elements to optimize to support more spatial and polarization channels. Unlike adopting a single 2D diffraction grating coupler in our approach, prior works also reported using grating coupler array for high-dimensional modes without polarization diversity \cite{seyedinnavadeh2024determining, butow2024generating, wu2023chip, sharma2025universal}. The effective field area is also limited which often results in reduced coupling efficiency. Hence, future designs must maximize channel dimensionality while minimizing coupling loss, potentially via poly-silicon or silicon nitride overlays atop gratings \cite{mak2018multi, zhou2025efficient, zhou2025adaptive}.

In our current experiment, thermal-optic phase shifters are employed to dynamically configure the on-chip SVD optical mesh and process the light field via a forward transmission process. The required reconfiguration time of our photonic processor is around a minute, mainly constrained by the feedback control electronics and the iterative algorithm. The operation speed of thermal effect on silicon typically falls within the $\mu s$ scale \cite{jacques2019optimization}, thus essentially limits the configuration speed of our photonic processor. While high-speed alternatives like PN-junction modulators are incompatible with interferometric meshes owing to prohibitive insertion loss\cite{han2025exploring, lu2025whispering}, thin film lithium niobate or lithium tantalate offer strong electro-optic tuning for sub-nanosecond switching \cite{boes2023lithium, wang2024lithium}. Real-time multidimensional beam processing will require co-designed optoelectronic integration enabling robust adaptability in various applications \cite{sacchi2025integrated, zanetto2023time}.

\section*{Methods}\label{sec4}
\noindent\textbf{Photonic chip design and fabrication:} The silicon photonic processor presented herein is constructed on a silicon-on-insulator (SOI) wafer platform, featuring a top silicon layer with a thickness of 220 nm and a buried oxide layer measuring 2 $\mu m$. The diffraction grating region is created through reactive-ion etching process with an etching depth of 70 nm. A full etch process is employed to form strip waveguides. Subsequently, the silicon devices are insulated with an oxide cladding, and thermal optical phase shifters are integrated using Titanium nitride (TiN) heaters. The fabrication of the silicon photonic chip is carried out at the Advanced Micro Foundry (AMF) Silicon Photonics Platform.

\vspace{1em}
\noindent\textbf{Photonic processor configuration and experimental setup:} The few-mode fibers in our experiment are provided by OFS (Optical Fiber Solutions, two-mode graded-index fiber with core diameter of 16 $\mu m$ and NA = 0.14, four-mode graded-index fiber with core diameter of 20 $\mu m$ and NA = 0.19). The fiber mode-selective photonic lanterns are provided Phoenix Photonics (3PLS-GI-15-MS and 6PLS-GI-15-MS). An external tunable continuous-wave laser source (Santec TSL-770) is used with a multichannel optical power meter (Santec MPM-210H and MPM-215). The integrated photonic processor is driven by a multichannel programmable power supply (TIME-TRANSFER T-MS128-12CV), which is automatically controlled by a personal computer with feedback signals from an infrared camera (ARTCAM-991SWIR), or a commercial polarimeter (Thorlabs, PAX1000IR2/M), or multichannel optical power meter. The driving currents of photonic processor are self-configured by particle-swarm optimization algorithm. For high-speed optical communication experiments, commercial high-speed thin-film lithium niobate Mach-Zehnder modulators (NOEIC MZ135-LN-110-C-H) are employed to generate 64-Gbaud NRZ-OOK and PAM-4 optical signals. The modulators are driven by an arbitrary waveform generator (AWG, Keysight 8199A) followed with RF amplifiers (SHF T850C). The optical signals are boosted by erbium-doped fiber amplifiers (EDFA, Amonics AEDFA-PA-35-B-FA; Phoenix Photonics, FMEDFA-15-3-1-1-2) and sent to the photodiodes (Coherent XPDV4121R). Real-time oscilloscope (Keysight UXR0702AP) is utilized to captured the high-speed signals.

\medskip
\noindent \textbf{Additional Information}
\noindent Supplementary information is available in the online version of the paper.

\medskip
\noindent \textbf{Acknowledgments}
\noindent Y.T. acknowledges the support from the National Natural Science Foundation of China (No. 62305277), Guangdong Science and Technology Department (No.2025B1212150003, No.2024A1515012438), and Nansha District Key Area S\&T Scheme (No. 2024ZD007).

\medskip
\noindent \textbf{Author contributions}
\noindent The experiments are conceived by Z.C., W.Z., J.H., and Y.T.; Z.C., H.C., W.Z., and K.L. designed the photonic integrated circuits; The experiments are conducted by Z.C., W.Z., H.C. W.T., M.Z., and X.P.; K.L. Y.C. and Y.Y., designed the printed circuit boards; All authors participated in writing and revising the manuscript. The project is under the supervision of Y.T. 

\medskip
\noindent \textbf{Conflict of interest} The authors declare no conflict of interest.

\medskip
\noindent \textbf{Data Availability} The data that support the findings of this study are included in the article and its supplementary information. Other data are available from the corresponding author upon request.

\section*{Reference}
\bibliographystyle{naturemag}
\bibliography{0_arxiv}

\end{document}